\documentclass[useAMS,usenatbib,article]{mnras}

\usepackage{longtable}
\usepackage[]{graphicx}
\usepackage{amsmath}
\usepackage{natbib}
\usepackage{soul}
\title[Explaining the reportedly over-massive black holes]
{Explaining the reportedly over-massive black holes in early-type galaxies with intermediate-scale discs}

\author[G.~A.~D. Savorgnan \& A.~W. Graham]
{\parbox{\textwidth}{
Giulia A.~D.~Savorgnan$^{1}$\thanks{E-mail: \texttt{gsavorgn@astro.swin.edu.au}},
Alister W.~Graham$^{1}$}\vspace{0.4cm}\\
\parbox{\textwidth}{
$^{1}$Centre for Astrophysics and Supercomputing, Swinburne University of Technology, Hawthorn, Victoria 3122, Australia.\\}}

\pagerange{\pageref{firstpage}--\pageref{lastpage}} \pubyear{2015}

\begin{document}

\maketitle

\label{firstpage}

\begin{abstract}
The classification ``early-type'' galaxy includes both elliptically- and lenticular-shaped galaxies. 
Theoretically, the spheroid-to-disc flux ratio of an early-type galaxy can assume any positive value,  
but in practice studies often consider only spheroid/disc decompositions 
in which the disc neatly dominates over the spheroid at large galaxy radii, 
creating an inner ``bulge'' as observed in most spiral galaxies. 
Here we show that decompositions in which the disc remains embedded within the spheroid, 
labelled by some as ``unphysical'',  
correctly reproduce both the photometric and kinematic properties of early-type galaxies 
with intermediate-scale discs. 
Intermediate-scale discs have often been confused with large-scale discs and incorrectly modelled as such; 
when this happens, the spheroid luminosity is considerably underestimated. 
This has recently led to some surprising conclusions, 
such as the claim that a number of galaxies with intermediate-scale discs (Mrk 1216, NGC 1277, NGC 1271, and NGC 1332) 
host a central black hole whose mass is abnormally large compared to expectations from the (underestimated) spheroid luminosity. 
We show that when these galaxies are correctly modelled, 
they no longer appear as extreme outliers in the (black hole mass)-(spheroid mass) diagram. 
This not only nullifies the need for invoking different evolutionary scenarios for these galaxies 
but it strengthens the significance of the observed (black hole mass)-(spheroid mass) correlation 
and confirms its importance as a fundamental ingredient for theoretical and semi-analytic models 
used to describe the coevolution of spheroids and their central supermassive black holes. 

\end{abstract}


\section{Introduction}
\label{sec:int}
\begin{figure}
\begin{center}
\includegraphics[width=\columnwidth]{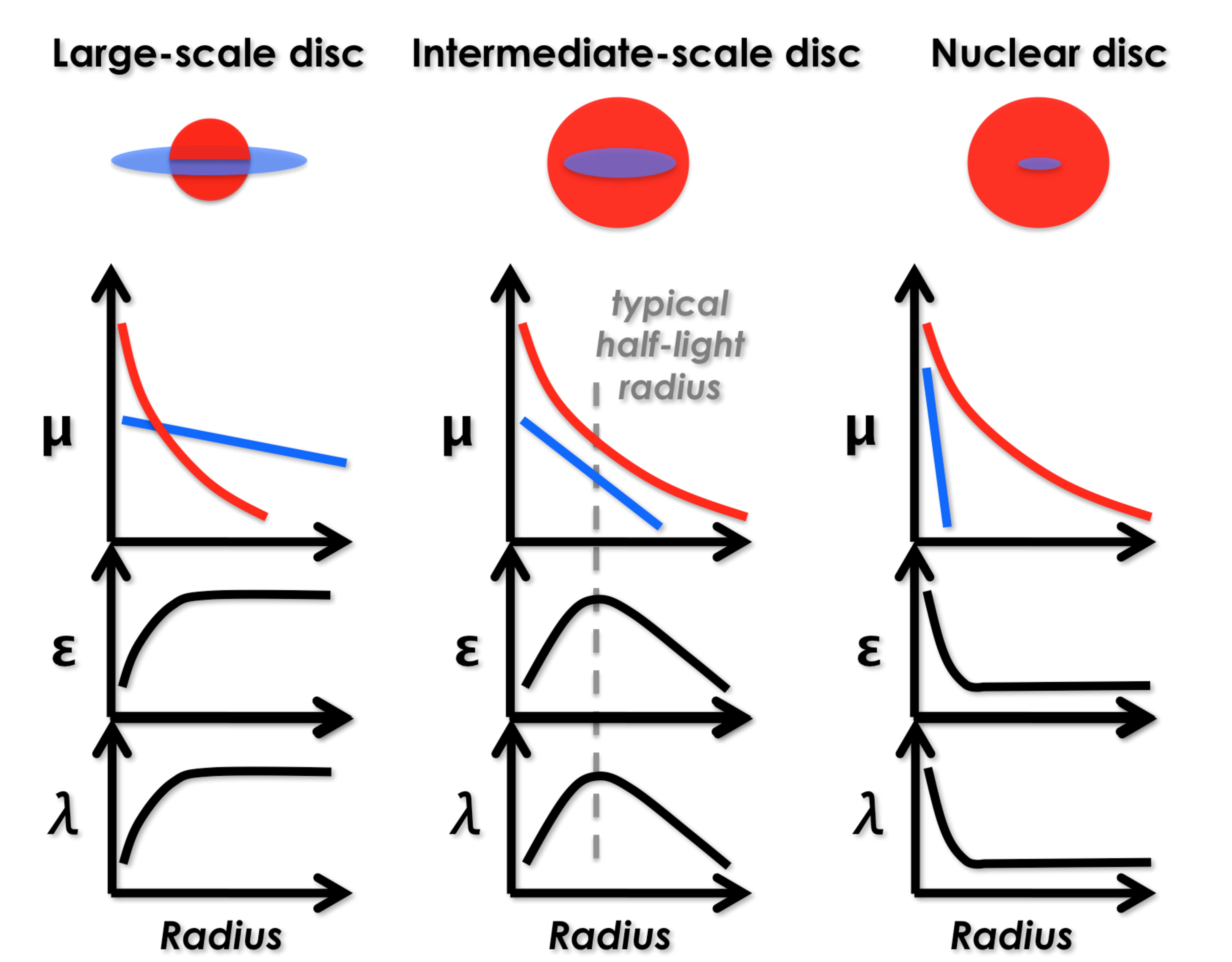}
\caption{Illustration of the spheroid/disc decomposition of the one-dimensional surface brightness profile, $\mu$, 
the ellipticity profile, $\epsilon$, and the specific angular momentum profile, $\lambda$,
for the three prototype early-type galaxy sub-classes. 
In the flux decompositions, the spheroid (or bulge) and the disc are shown with the red and blue color, respectively. 
The left panel shows a disc-dominated central fast rotator (lenticular galaxy), composed of a bulge encased in a large-scale disc. 
The right panel displays a spheroid-dominated slow rotator (elliptical) with (an optional) nuclear stellar disc. 
The middle panel presents a spheroid-dominated central fast rotator with an intermediate-sized disc embedded in the spheroid. }
\label{fig:model}
\end{center}
\end{figure}

The awareness that \emph{many} early-type galaxies contain previously over-looked stellar discs dates back half a century 
\citep{liller1966,stromstrom1978,michard1984,djorgovski1985,jedrzejewski1987,BenderMoellenhoff1987,
carter1987,capaccioli1987,capaccioli1988}. 
It is well known that 
the identification of a stellar disc in an early-type galaxy, particularly when based on the galaxy's photometric properties, 
is subject to inclination effects. 
As predicted by \cite{carter1987}, this problem is largely overcome with kinematic analyses 
(e.g.~\citealt{franx1989,nieto1991,rixwhite1992,cinzanovandermarel1993,donofrio1995,graham1998fornax}, 
and the ATLAS$^{\rm 3D}$ survey, \citealt{cappellari2011}), 
which allow one to determine the presence of a rotationally-supported component 
in a way nearly insensitive to projection effects \citep{mcelroy1983,cappellari2007,emsellem2007}. 
Yet, identifying the radial extent of an early-type galaxy's disc with respect to the spheroidal component can still be subtle. 
Studying both the surface brightness profiles and the ellipticity profiles 
of early-type galaxies in the Virgo cluster -- including those with elliptical (E), spindle and lenticular (S0) isophotes -- 
\cite{liller1966} drew attention to the observation that many of the galaxies displayed 
``characteristics intermediate between those of type E and type S0'', 
and she classified them as ``ES'' galaxies.  
Building on this and other investigations of ellipticity profiles (e.g.~\citealt{stromstrom1978,ditullio1979}), 
\cite{michard1984} used the classification ``S0-like'' for these early-type galaxies with humped ellipticity profiles, 
dominated by a somewhat edge-on disc at intermediate radii.  
\cite{nieto1988} identified two dozen such spheroid-dominated early-type galaxies, 
whose discs do not prevail at large radii, 
and referred to them as ``disk-ellipticals'' (or ``disky ellipticals'', \citealt{simienmichard1990}).   
However, as noted by \cite{nieto1988}, unless the orientation of the disc is favourable 
(i.e.~somewhat edge-on), it can be missed.  
The same is true when searching for pointy isophotes that are shaped by the combination of the spheroid and a near edge-on disc 
(e.g.~\citealt{carter1978,carter1987,jedrzejewski1987,ebneter1987,BenderMoellenhoff1987,bender1988,bijaoui1989}). \\
Today, most early-type galaxies are classified as ``fast rotators'' 
\citep{atlas3dIII,scott2014}, 
that is, they are rapidly rotating within their half-light radius. 
The exact definition of a fast rotator can be found in \cite{emsellem2007}, 
although the most recent literature (e.g.~\citealt{arnold2011n3115,romanowskyfall2012,arnold2014}) 
prefers the use of the term ``\emph{central} fast rotator'' 
to emphasize the fact that this classification pertains to the kinematic properties of a galaxy only within its half-light radius.
Thanks to their more extended kinematic maps, 
\cite{arnold2014} revealed that some of the central fast rotators continue to be fast rotating at large radii, 
whereas other central fast rotators become slow rotating in their outer regions\footnote{As pointed out by \cite{cappellari2011}, 
while all of the disky ellipticals from \cite{Bender1994} are fast rotators, 
the complement is not true because weak discs only impact the isophotal shape if the discs have orientations close to edge-on, 
whereas their rotational signature can still be detected when they have a near face-on orientation.  
Of course if a disc is face-on, then the galaxy will not be classified as a fast rotator. }.
Unfortunately, such extended kinematic maps are not yet available for large numbers of galaxies in the local Universe. 
Nevertheless, the ellipticity profile of a galaxy's isophotes can help identify the extent of a stellar disc in an early-type galaxy. \\
In general, stellar discs are intrinsically flat and close to circular (e.g.~\citealt{Andersen2001,AndersenBershady2002}); 
their apparent ellipticity, dictated by their inclination to our line of sight, is fixed. 
Spheroids are often rounder than the observed projection on the sky of their associated discs, 
thus their average ellipticity is often lower than that of their disc. 
An ellipticity profile that increases with radius can be ascribed to an inclined disc that becomes progressively more important at large radii, 
whereas a radial decrease of ellipticity signifies the opposite case. 
This approach can be taken to the next level by inspecting the isophotes for discy structures 
(e.g.~\citealt{carter1978,carter1987,capaccioli1987,jedrzejewski1987,BenderMoellenhoff1987}) 
and checking the velocity line profiles for asymmetry 
(e.g.~\citealt{franxillingworth1988ic1459,bender1990,rixwhite1992,scorzabender1995}, and references therein; \citealt{scorza1998}). \\
Building on the investigations in works such as \cite{liller1966}, \cite{Jedrzejewski1987INPROCEEDINGS} and \cite{rixwhite1990}, 
the toy model shown in Figure \ref{fig:model} illustrates the typical ellipticity profile 
($\epsilon = 1 - b/a$, where $b/a$ is the ratio of minor-to-major axis length) 
and the specific angular momentum profile 
($\lambda = \langle R |V| \rangle / \langle R \sqrt{V^2 + \sigma^2} \rangle$, 
where $R$ is the semimajor-axis radius, $V$ is the mean velocity and $\sigma$ is the velocity dispersion, \citealt{emsellem2007}) 
of: 
(i) a lenticular galaxy, 
comprised of a large-scale disc which dominates the light at large radii over a relatively smaller encased bulge,  
i.e.~a disc-dominated central fast rotator that continues to be fast rotating beyond one half-light radius; 
(ii) a ``discy elliptical'' galaxy \citep{michard1984,nieto1988} 
composed of an intermediate-scale disc embedded in a relatively larger spheroid which dominates the light at large radii,
i.e.~a spheroid-dominated central fast rotator that becomes slow rotating beyond $1-2$ half-light radii; and  
(iii) an elliptical galaxy with an additional nuclear stellar disc, 
i.e.~a (spheroid-dominated) slow rotator. 
This sequence is analogue to that illustrated in Figure 2 of \cite{cappellari2011kmdr}, 
although here we emphasize the correspondence between the spheroid/disc decomposition of the surface brightness profile 
and the ``shape'' of the ellipticity profile (assuming that the disc inclination is not close to face-on) 
and also the specific angular momentum profiles. \\
While some recent studies have correctly distinguished between large- and intermediate-scale discs, 
and modelled them accordingly (e.g.~\citealt{kormendybender2012,krajnovic2013}), 
intermediate-scale discs have been missed by many galaxy modellers of late, 
who have labelled as ``unphysical'' \citep{allen2006} those spheroid/disc decompositions 
in which the disc does not dominate over the spheroid at large radii 
as is observed with spiral galaxies. 
This has led to the rejection of many early-type galaxy decompositions 
similar to that illustrated in the top middle panel of Figure \ref{fig:model}. 
Unsurprisingly, studies affected by this bias have not obtained spheroid/disc decompositions with a spheroid-to-total ratio larger than $0.6 - 0.8$ 
(e.g.~\citealt{gadotti2008,head2014,querejeta2015,mendezabreu2015}). \\
As mentioned before, an isophotal analysis allows one to identify the presence and the radial extent of a disc in an early-type galaxy 
only when the disc has a certain level of inclination. 
On the other hand, a kinematic analysis has the advantage of being virtually insensitive to inclination effects, 
but cannot help one determine the radial extent of a disc if the kinematic data are limited within one half-light radius. 
Therefore, the best results are obtained when photometry and kinematics are combined together. \\
In this paper we focus on the increasingly overlooked occurence of intermediate-scale discs in galaxies with directly measured black hole masses. 
We report on the photometric and kinematical signatures of these intermediate-sized stellar discs,  
and the impact they have on the (black hole mass)-to-(spheroid stellar mass) ratio 
which is used to constrain galaxy evolution models. 
In Section \ref{sec:gal} we present a detailed photometric analysis of three galaxies with intermediate-scale discs (Mrk 1216, NGC 1332, and NGC 3115) 
and we briefly describe another five galaxies with intermediate-scale discs (NGC 821, NGC 1271, NGC 1277, NGC 3377, and NGC 4697) 
already modelled by us elsewhere in the literature. 
We compare our photometric analysis with the kinematical information available from the literature, 
and explain the differences between our galaxy models and past decompositions. 
In Section \ref{sec:mm} we explore the important implications this has for the (black hole mass)-(spheroid stellar mass) diagram. 
Finally, in Section \ref{sec:disc} we briefly discuss our results in terms of galaxy evolution.

\section{Intermediate-scale disc galaxies}
\label{sec:gal}
Three examples of galaxies with intermediate-scale discs are Mrk 1216, NGC 1332, and NGC 3115. 
In the following Section, we present a photometric analysis of these three galaxies, 
and we compare our results with the kinematical analysis available from the literature for Mrk 1216 and NGC 3115. 
For the galaxies NGC 1332 and NGC 3115, we used $3.6~\rm \mu m$ images obtained with the InfraRed Array Camera (IRAC) 
onboard the \emph{Spitzer Space Telescope}. 
For the galaxy Mrk 1216, we used an archived Hubble Space Telescope (\emph{HST}) image  
taken with the Wide Field Camera 3 (WFC3) and the near-infrared \emph{F160W} filter ($H$-band). 
Our galaxy decomposition technique is extensively described in \cite{paperI}.
Briefly, the galaxy images were background-subtracted, and masks for contaminating sources were created. 
The one-dimensional Point Spread Function (PSF) was characterized using a Gaussian profile for the \emph{HST} observation 
and a \cite{moffat1969} profile for the \emph{Spitzer} observations.
We performed an isophotal analysis of the galaxies using the IRAF\footnote{IRAF 
is the Image Reduction and Analysis Facility, distributed by the National Optical Astronomy Observatory, 
which is operated by the Association of Universities for Research in Astronomy (AURA) 
under cooperative agreement with the National Science Foundation.} task {\tt ellipse}\footnote{Our analysis 
was performed before {\tt isofit} \citep{ciambur2015} was conceived or available. 
After {\tt isofit} was recently developed and implemented in IRAF, 
we employed it to re-extract the surface brightness profiles of the galaxies NGC 1332 and NGC 3115. 
We then repeated the analysis and checked that this change does not significantly alter our results. 
In fact, although {\tt isofit} provides a more accurate description of the isophotes in the presence of an inclined disc, 
the discs of NGC 1332 and NGC 3115 are relatively faint compared to the spheroidal components, 
therefore the differences between the light profile obtained with {\tt ellipse} and that obtained with {\tt isofit} 
are small for these two galaxies. } 
\citep{taskellipse}. 
The galaxy isophotes were modelled with a series of concentric ellipses, 
allowing the ellipticity, the position angle and the amplitude of the fourth harmonic to vary with radius.  
The decomposition of the surface brightness profiles was performed with software written by G. Savorgnan 
and described in \cite{paperI}.
We modelled the light profiles with a combination of PSF-convolved analytic functions, 
using one function per galaxy component.

\subsection{NGC 3115}
\begin{figure}
\begin{center}
\includegraphics[width=0.49\columnwidth]{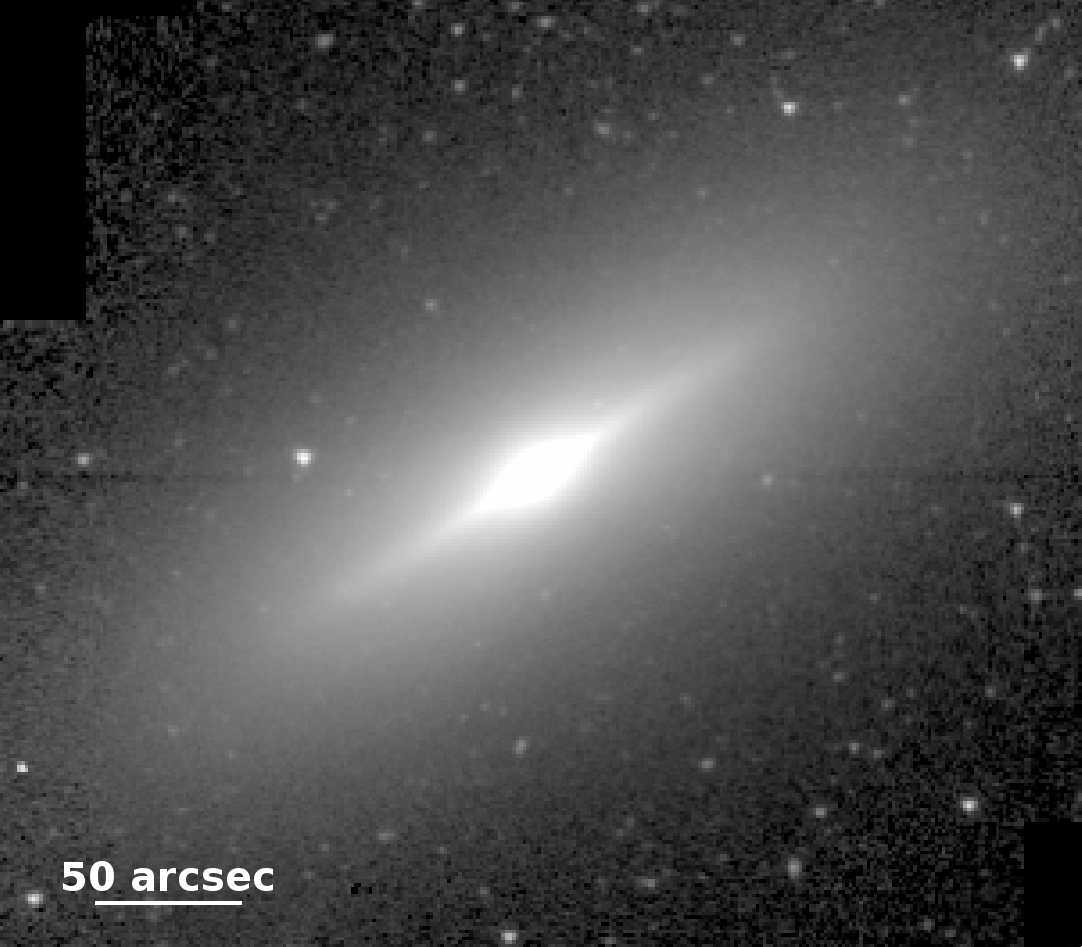}
\includegraphics[width=0.49\columnwidth]{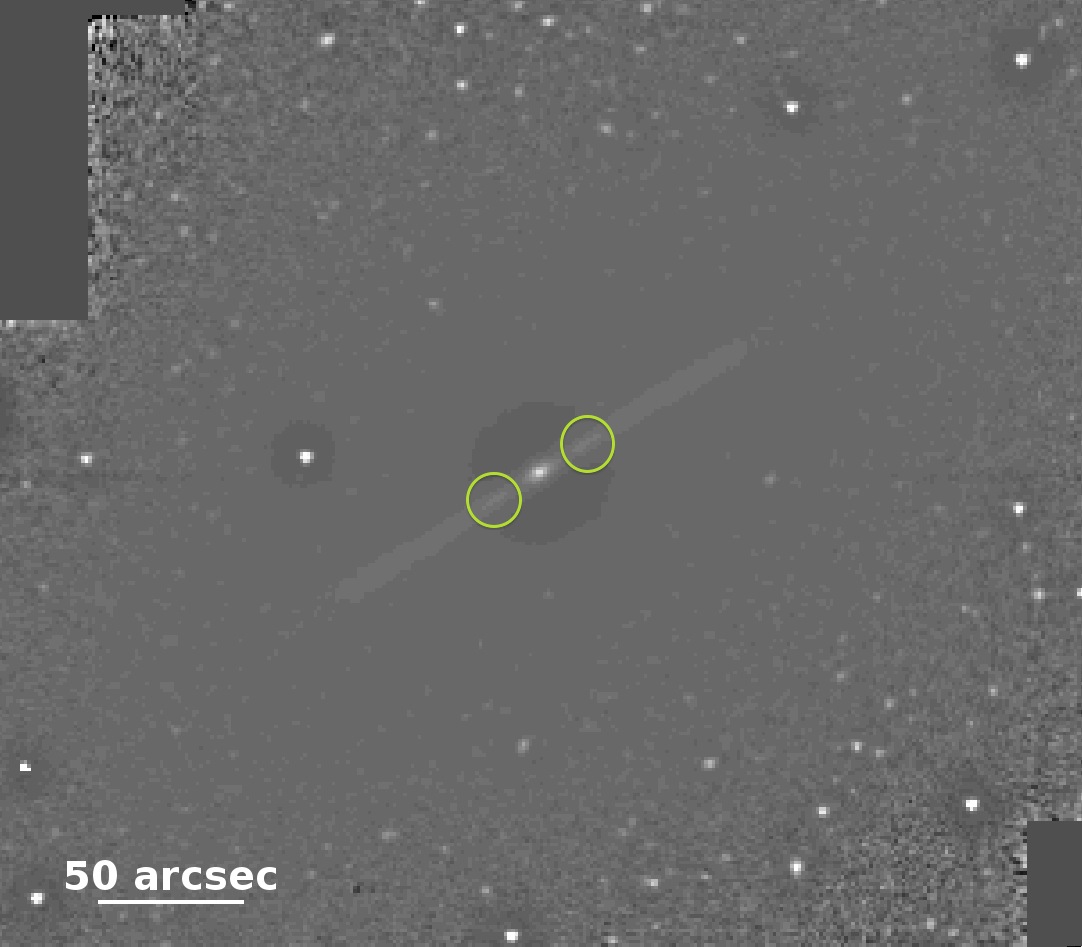} \\
\includegraphics[width=1.05\columnwidth]{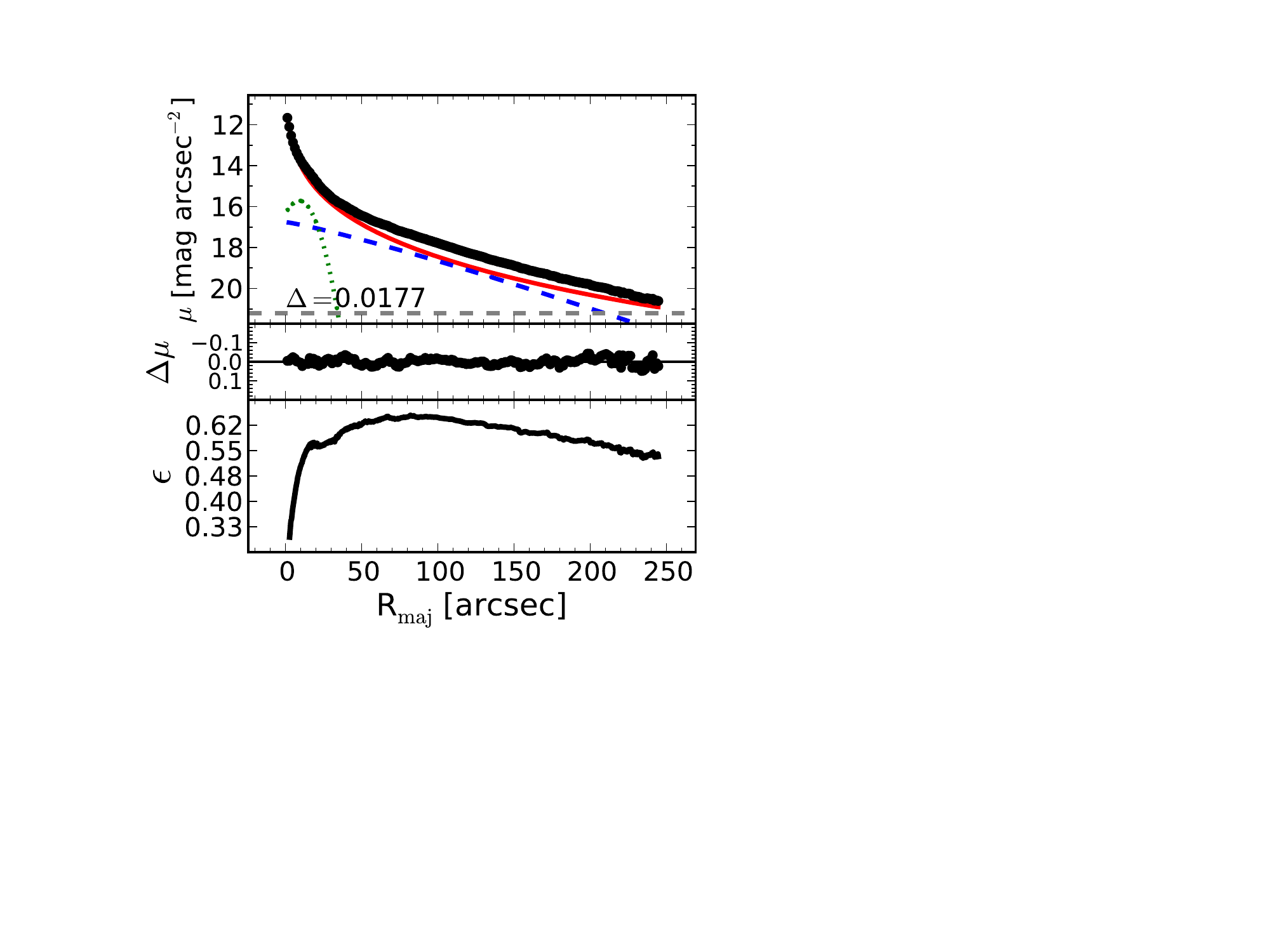}
\caption{NGC 3115. 
The top panels are the \emph{Spitzer}/IRAC $3.6~\rm \mu m$ image (left) and its unsharp mask (right), 
obtained by dividing the image by a Gaussian-smoothed version of itself. 
In the unsharp mask, the green circles indicate the position of the two brighter spots associated to the edge-on nuclear ring. 
The bottom plots display the best-fit model of the surface brightness profile, $\mu$, 
and the ellipticity profile, $\epsilon$, 
along the major-axis, $R_{\rm maj}$. 
The black points are the observed data, which extend out to five galaxy half-light radii ($\sim 5 \times 50~\rm arcsec$). 
The color lines represent the individual (PSF-convolved) model components: 
red solid = S\'ersic (spheroid), blue dashed = S\'ersic (disc), green dotted = Gaussian ring. 
The residual profile (data $-$ model) is shown as $\Delta \mu$. 
The horizontal gray dashed line corresponds to an intensity equal to three times the root mean square of the sky background fluctuations. 
$\Delta$ denotes the root mean square scatter of the fit in units of $\rm mag~arcsec^{-2}$. }
\label{fig:n3115}
\end{center}
\end{figure}

The presence of a disc in the central fast rotator NGC 3115 
(e.g.~\citealt{strom1977,nieto1988,scorzabender1995}) 
is obvious due to its edge-on orientation (Figure \ref{fig:n3115}). 
Less obvious is the radial extent of this disc if one only relies on a visual inspection of the galaxy image. 
The ellipticity profile (Figure \ref{fig:n3115}) is consistent with the presence of an intermediate-scale disc. 
Moreover, the kinematics of NGC 3115 \citep{arnold2011n3115} also disprove the presence of a large-scale disc, 
because the galaxy is rapidly rotating only within two galaxy half-light radii ($\sim 2 \times 50~\rm arcsec$), 
and the rotation significantly drops at larger radii.  
The unsharp mask of NGC 3115 (Figure \ref{fig:n3115}) betrays the presence of a faint edge-on nuclear ring, 
which can also be spotted as a small peak in the ellipticity profile 
(at semi-major axis length $R_{\rm maj} \sim 15~\rm arcsec$). 
Such rings are common in early-type galaxies (e.g.~\citealt{michardmarchal1993}).
The spheroidal component of NGC 3115 is well described with a \cite{sersic1963} profile.
The highly inclined intermediate-scale disc is better fit with an $n<1$ S\'ersic profile 
(the S\'ersic index $n$ regulates the curvature of the S\'ersic profile) 
rather than with an exponential function, 
as explained by \cite{pastrav2013a}. 
The nuclear ring is modelled with a Gaussian function. \\
In comparison, \cite{lasker2014data} fit NGC 3115 with a bulge + disc + envelope, 
and measured a bulge half-light radius of $3.9~\rm arcsec$ and a bulge-to-total ratio of $0.12$. 
We describe this galaxy using a spheroid + intermediate-scale disc + nuclear ring, 
and obtain a spheroid half-light radius of $43.6~\rm arcsec$ and a spheroid-to-total ratio of $0.85$. 
We have used both kinematical information and ellipticity profiles, 
together with the surface brightness profile, 
to obtain a physically consistent and meaningful model.

\subsection{NGC 1332}
\begin{figure}
\begin{center}
\includegraphics[width=0.49\columnwidth]{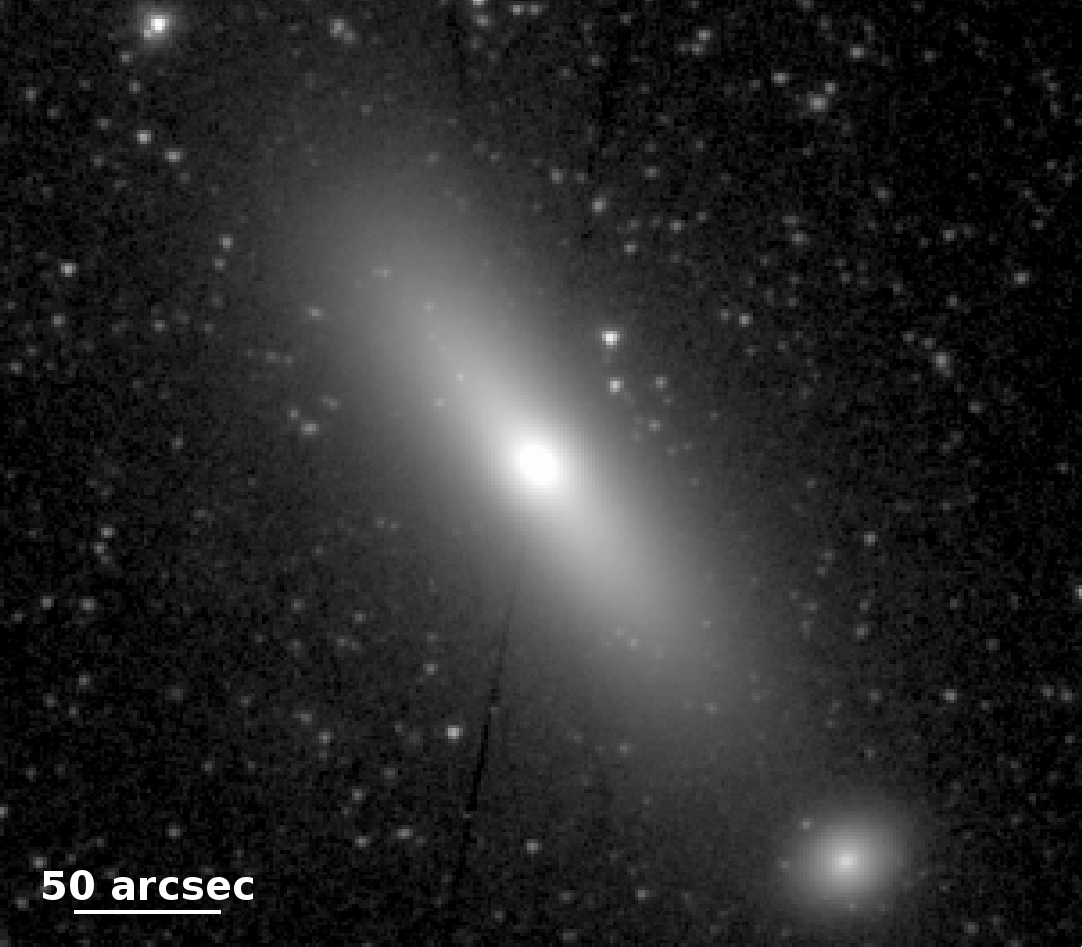}
\includegraphics[width=0.49\columnwidth]{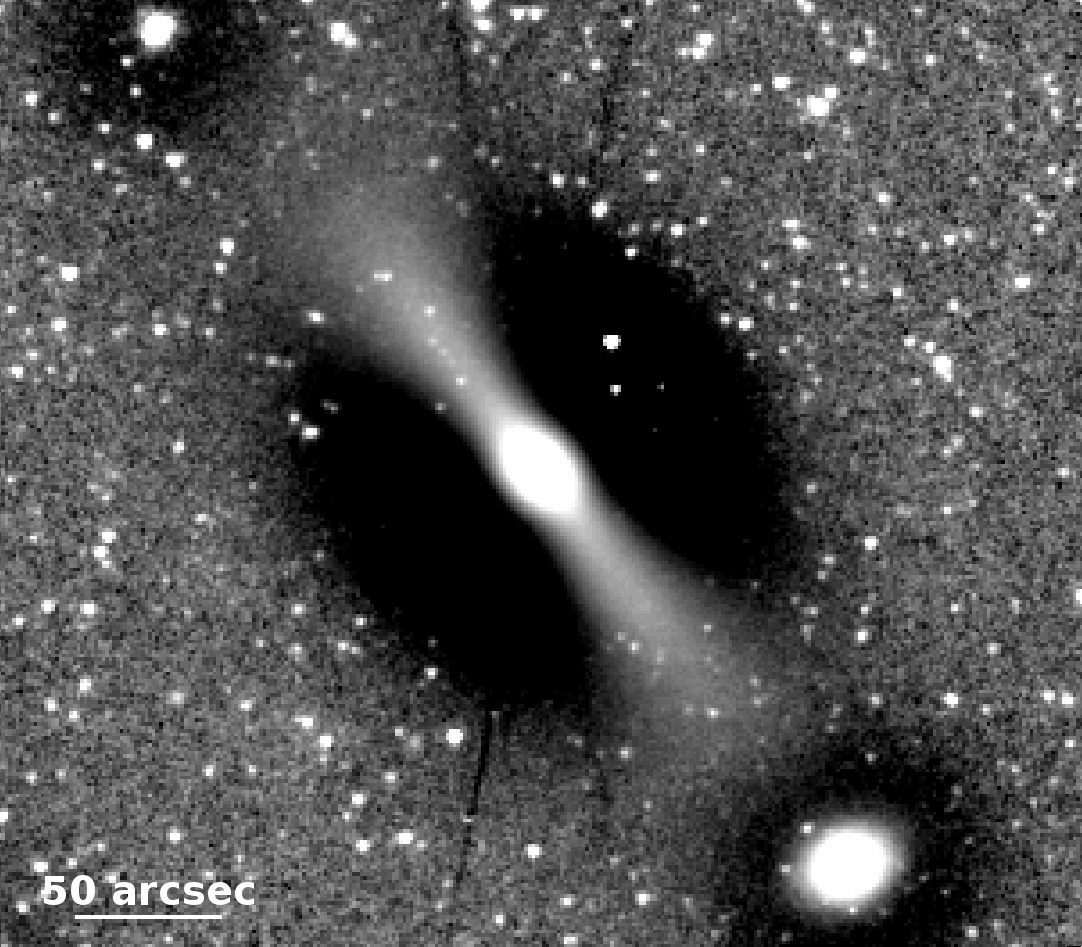} \\
\includegraphics[width=1.03\columnwidth]{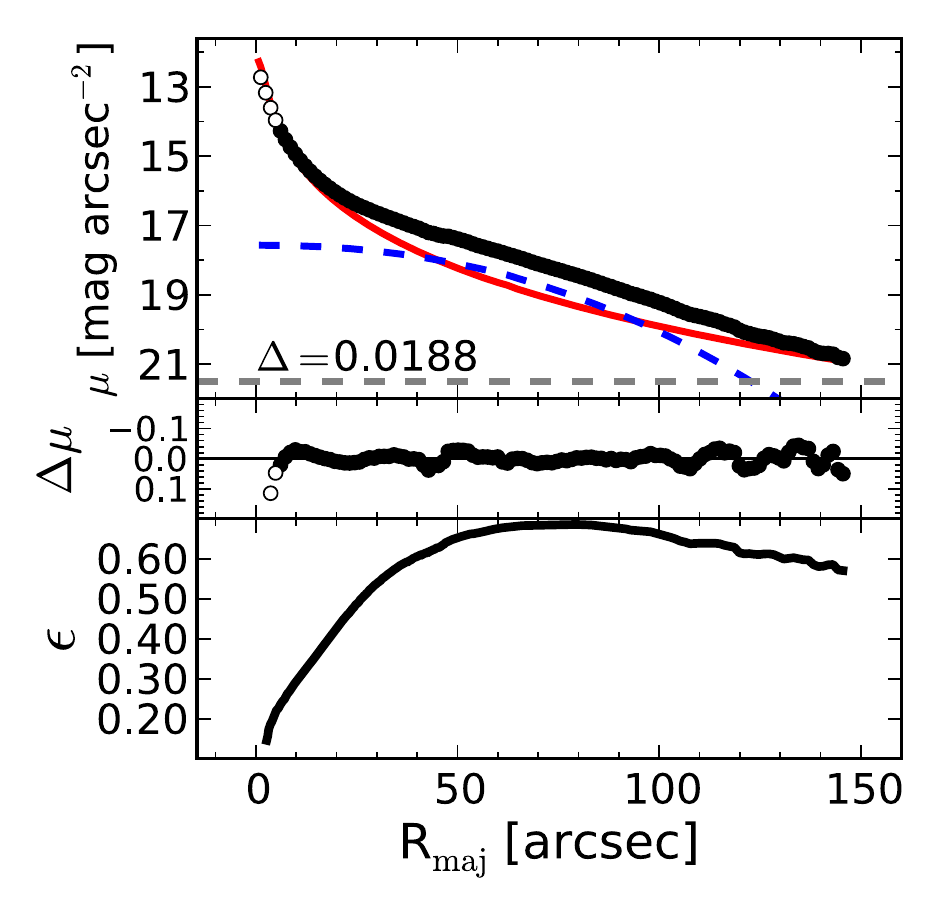}
\caption{NGC 1332.
Similar to Figure \ref{fig:n3115}. 
The surface brightness profile extends out to seven galaxy half-light radii ($\sim 7 \times 20~\rm arcsec$). 
The empty points are data excluded from the fit. 
}
\label{fig:n1332}
\end{center}
\end{figure}

The morphology of NGC 1332 (Figure \ref{fig:n1332}) is very similar to that of NGC 3115, 
with the ellipticity profile indicating the presence of an intermediate-scale disc, 
although in this case no nuclear component is evident. 
We were not able to find any extended kinematic profile or map 
for this galaxy in the literature. 
The data within the innermost $6~\rm arcsec$ were excluded from the fit 
because, according to our galaxy decomposition, they are possibly affected by the presence of a partially depleted core.
The surface brightness profile of NGC 1332 is well described with a S\'ersic-spheroid plus
an $n<1$ S\'ersic-disc. 
Our galaxy decomposition suggests that NGC 1332 is a spheroid-dominated galaxy, 
with a spheroid-to-total ratio of $0.95$. \\
\cite{rusli2011} did not identify the restricted extent of the intermediate-scale disc, 
as revealed by the ellipticity profile, 
and proposed a model featuring a S\'ersic-bulge and a large-scale exponential-disc, 
with a spheroid-to-total ratio of $0.43$.
Based on their bulge/disc decomposition, they concluded that NGC 1332 is a disc-dominated lenticular galaxy 
which is displaced from the (black hole mass)-(spheroid luminosity) correlation of \cite{marconihunt2003} 
by an order of magnitude along the black hole mass direction. 
However, in Section \ref{sec:mm} we show that, according to our decomposition, 
NGC 1332 lies within the $1\sigma$ scatter about the (black hole mass)-(spheroid stellar mass) correlation 
for early-type galaxies. 
We also note that the majority of galaxies with an elevated stellar velocity dispersion ($\sigma > 270~\rm km~s^{-1}$) 
are core-S\'ersic galaxies \citep{graham2003coresersicmodel,ferrarese2006acsvcs,dullograham2014cores}, 
i.e.~they have a partially depleted core which has been identified from high-resolution photometric data. 
NGC 1332 has $\sigma = 320~\rm km~s^{-1}$, but, 
based on their decomposition of \emph{HST} imaging, \cite{rusli2011} did not find a core in this galaxy. 
However, our galaxy decomposition (Figure \ref{fig:n1332}) suggests that NGC 1332 is in fact a core-S\'ersic galaxy. 
Since we did not use high-resolution photometric data, 
we refrain from a firm conclusion, 
but we caution that a re-analysis of the \emph{HST} data -- by taking into account the correct radial extent of the intermediate-scale disc --
may indeed reveal the presence of a depleted core in this galaxy.

\subsection{Mrk 1216}
\begin{figure}
\begin{center}
\includegraphics[width=0.49\columnwidth]{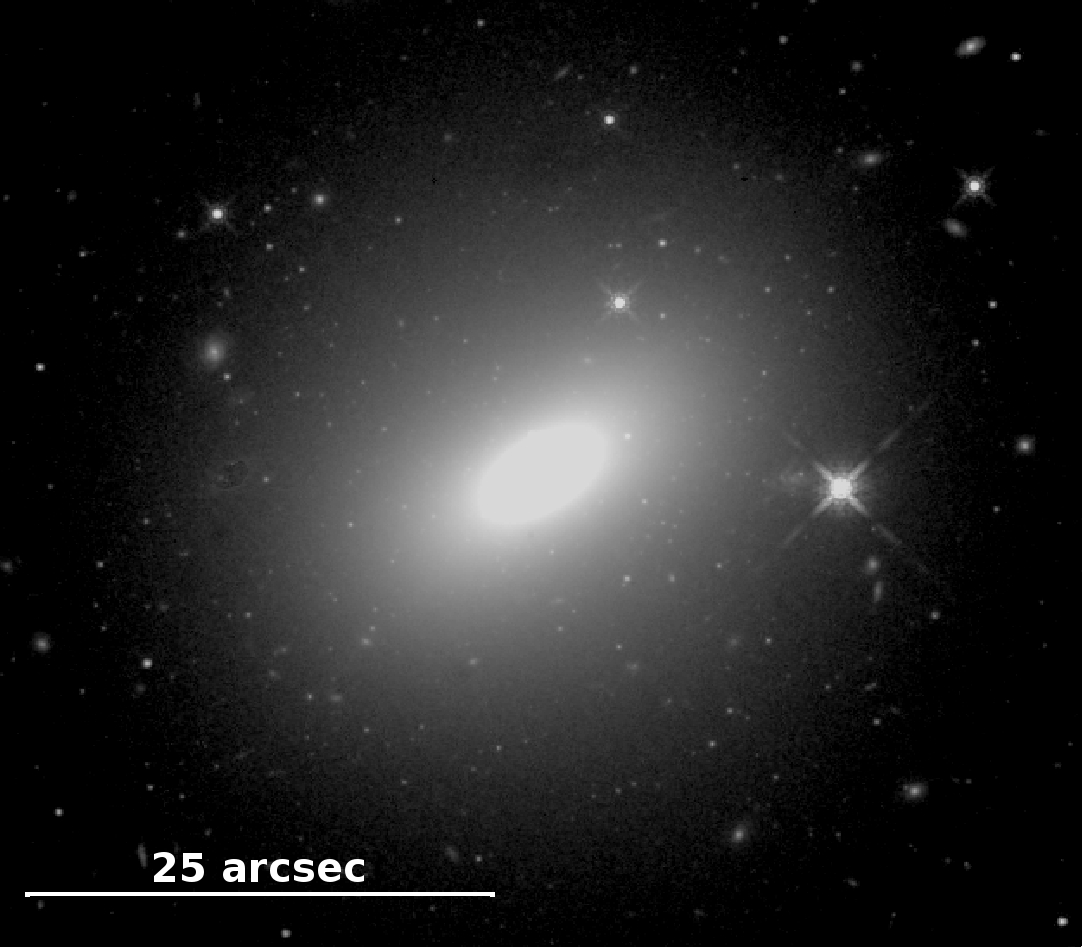}
\includegraphics[width=0.49\columnwidth]{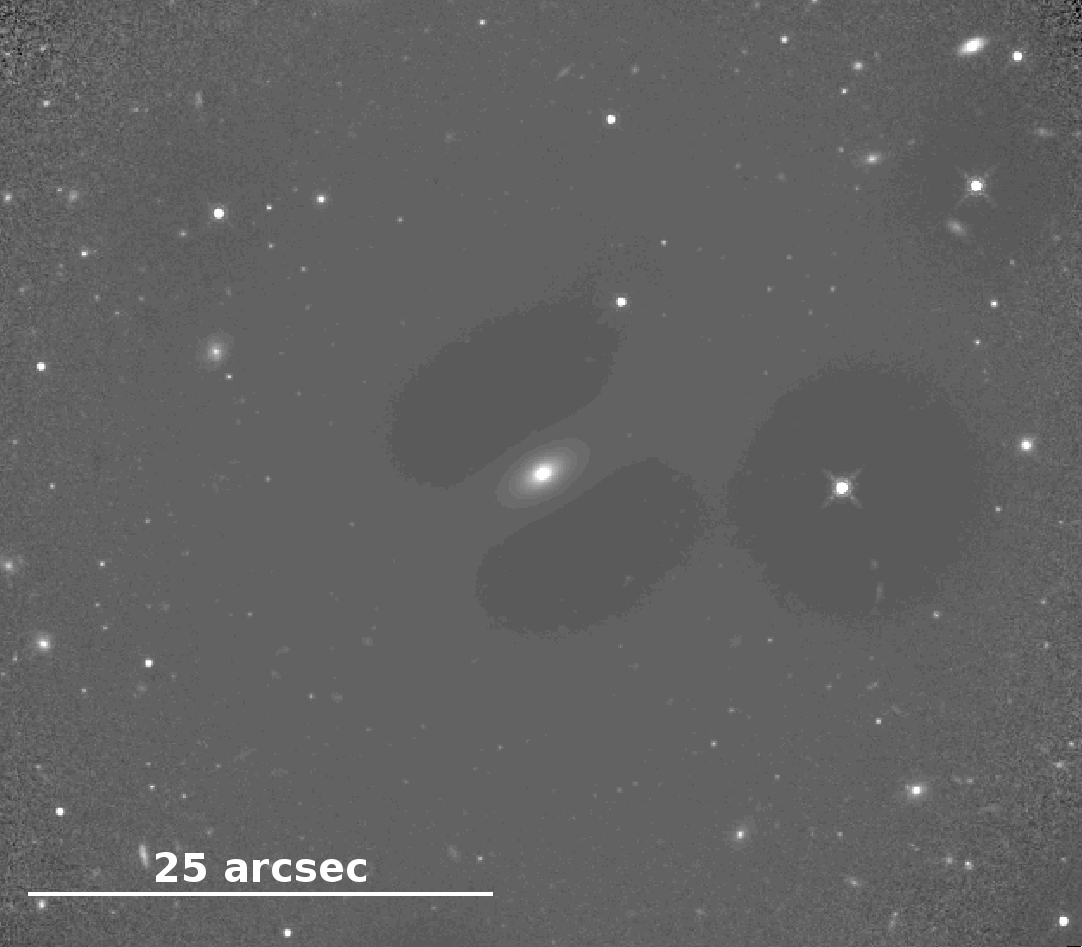} \\
\includegraphics[width=1.05\columnwidth]{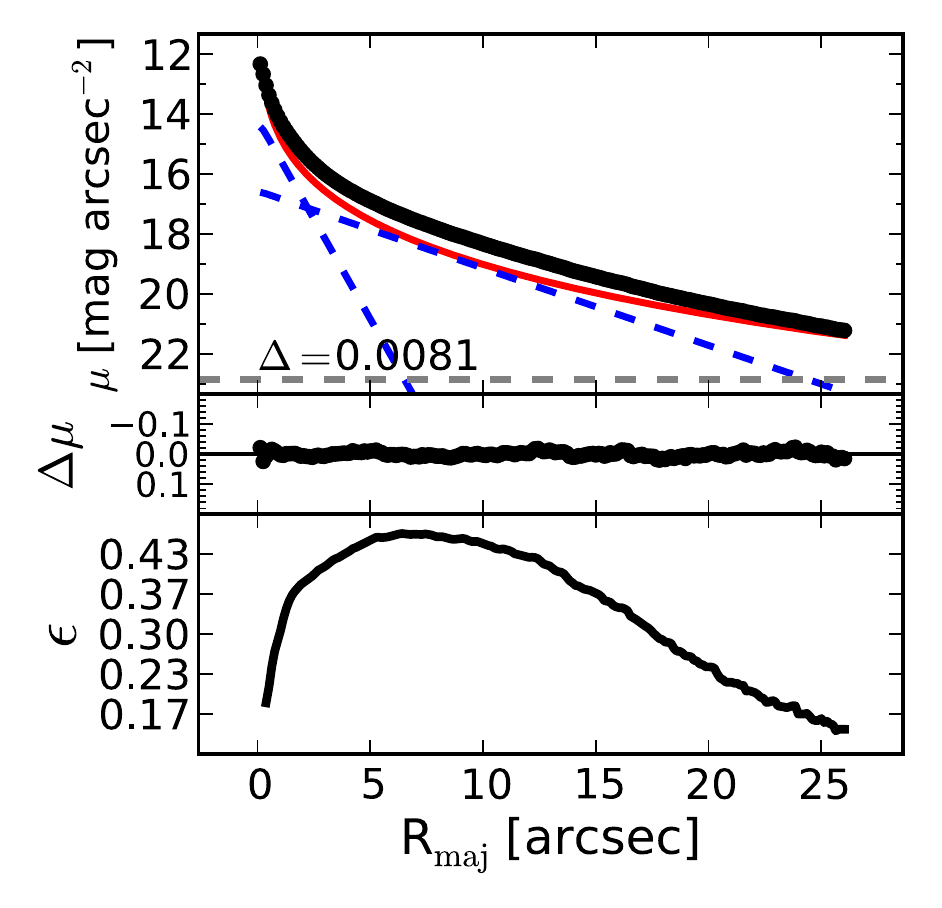}
\caption{Mrk 1216. 
Similar to Figure \ref{fig:n3115}. 
The top panels are the \emph{HST}/WFC3 \emph{F160W} image (left) and its unsharp mask (right).
The surface brightness profile extends out to five galaxy half-light radii ($\sim 5 \times 5~\rm arcsec$). 
The color lines represent the individual (PSF-convolved) model components:
red solid = S\'ersic (spheroid), blue dashed = exponential (nuclear and intermediate-scale disc). 
}
\label{fig:m1216}
\end{center}
\end{figure}

Although the disc in the central fast rotator Mrk 1216 is not immediately apparent from the image (Figure \ref{fig:m1216}), 
the velocity map \citep{yildirim2015} reveals the presence of a fast rotating component 
within three galaxy half-light radii ($\sim 3 \times 5~\rm arcsec$). 
The ellipticity profile (Figure \ref{fig:m1216}), 
which extends out to five half-light radii, indicates the presence of an intermediate-scale disc. 
In addition, a nuclear disc is identified from the change in slope of the ellipticity profile ($R_{\rm maj} \sim 1 - 2 \rm~arcsec$), 
from the unsharp mask, 
and from a clear feature in the $B4$ fourth harmonic profile (not shown here). 
We modelled the surface brightness profile of Mrk 1216 (Figure \ref{fig:m1216}) with a S\'ersic-spheroid, 
an intermediate-sized exponential-disc, and a nuclear exponential-disc. 

\subsection{Other galaxies}
Our models with an intermediate-sized disc embedded within a larger spheroidal component, 
plus an additional nuclear component when one is present, 
match the observed light distribution, and explain both the extended kinematic maps (when available, \citealt{arnold2014}) and the ellipticity profiles, 
of five additional galaxies for which a direct measurement of their central supermassive black hole mass is available: 
NGC 821; NGC 1271; NGC 1277; NGC 3377; and NGC 4697. 
Our isophotal analysis and galaxy decompositions for NGC 1271 and NGC 1277 will be presented in 
Graham, Savorgnan \& Ciambur (\emph{in prep.}) and \cite{graham2015n1277}, respectively, 
while the galaxies NGC 821, NGC 3377 and NGC 4697 have been analysed in \cite{paperI}. 

\subsubsection{NGC 1271}
\cite{walsh2015} explored a three-component decomposition for the central fast rotator NGC 1271 
and identified the galaxy bulge with the innermost of the three components, 
having a half-light radius of $0.61~\rm arcsec$ and a bulge-to-total flux ratio of $0.23$; 
our model features a spheroid + intermediate-scale disc, 
with a spheroid half-light radius of $3.3~\rm arcsec$ and a spheroid-to-total flux ratio of $0.67$. 

\subsubsection{NGC 1277}
\cite{vandenbosch2012} proposed a model for the central fast rotator NGC 1277 with a bulge + disc + nuclear source + envelope, 
which gives a bulge half-light radius of $0.9~\rm arcsec$ and a bulge-to-total flux ratio of $0.24$; 
our model consists of a spheroid + intermediate-scale disc + nuclear component, 
and produces a spheroid half-light radius of $6.0~\rm arcsec$ and a spheroid-to-total flux ratio of $0.79$. 

\subsubsection{NGC 3377}
\cite{lasker2014data} modelled the central fast rotator NGC 3377 
(e.g.~\citealt{Jedrzejewski1987INPROCEEDINGS,scorzabender1995}) 
with a bulge + nuclear disc + disc + envelope, 
and obtained a bulge half-light radius of $10.1~\rm arcsec$ and a bulge-to-total flux ratio of $0.35$; 
our model with a spheroid + intermediate-scale disc + nuclear disc 
returns a spheroid half-light radius of $61.8~\rm arcsec$ and a spheroid-to-total flux ratio of $0.94$. 

\subsubsection{NGC 821}
\cite{lasker2014data} decomposed the central fast rotator NGC 821 into a bulge + disc + envelope, 
and measured a bulge half-light radius of $3.8~\rm arcsec$ and a bulge-to-total flux ratio of $0.19$; 
our decomposition consists of a spheroid + intermediate-scale disc, 
with a spheroid half-light radius of $36.5~\rm arcsec$ and a spheroid-to-total flux ratio of $0.79$. 

\subsubsection{NGC 4697}
While NGC 4697 (e.g.~\citealt{carter1987,Jedrzejewski1987n720n1052n4697,davies1981}) 
was explicitly referred to as a ``fast rotator'' by \cite{capaccioli1987} and \cite{petrou1981}, 
it is only a central fast rotator and it represents an ``extreme'' case. 
\cite{lasker2014data} fit this galaxy with a bulge + nuclear source + disc + envelope, 
and obtained a bulge half-light radius of $6.3~\rm arcsec$ and a bulge-to-total flux ratio of $0.08$; 
we described NGC 4697 using a spheroid + intermediate-scale disc + nuclear disc model, 
and measured a spheroid half-light radius of $239.3~\rm arcsec$ and a spheroid-to-total flux ratio of $0.89$. \\

Past models that ``forcedly'' described intermediate-scale disc galaxies using an inner bulge 
encased within a large-scale disc 
commonly required the addition of an extended envelope or halo to account for the outer portion of the spheroid. 
Such three-component models (bulge + disc + envelope) typically reduce the spheroid luminosity by a factor of $3-4$, 
and underestimate the size of the spheroid by a factor of $6-10$, 
although more ``extreme'' cases can be found.

\section{The black hole -- spheroid correlation}
\label{sec:mm}
\begin{figure}
\begin{center}
\includegraphics[width=\columnwidth]{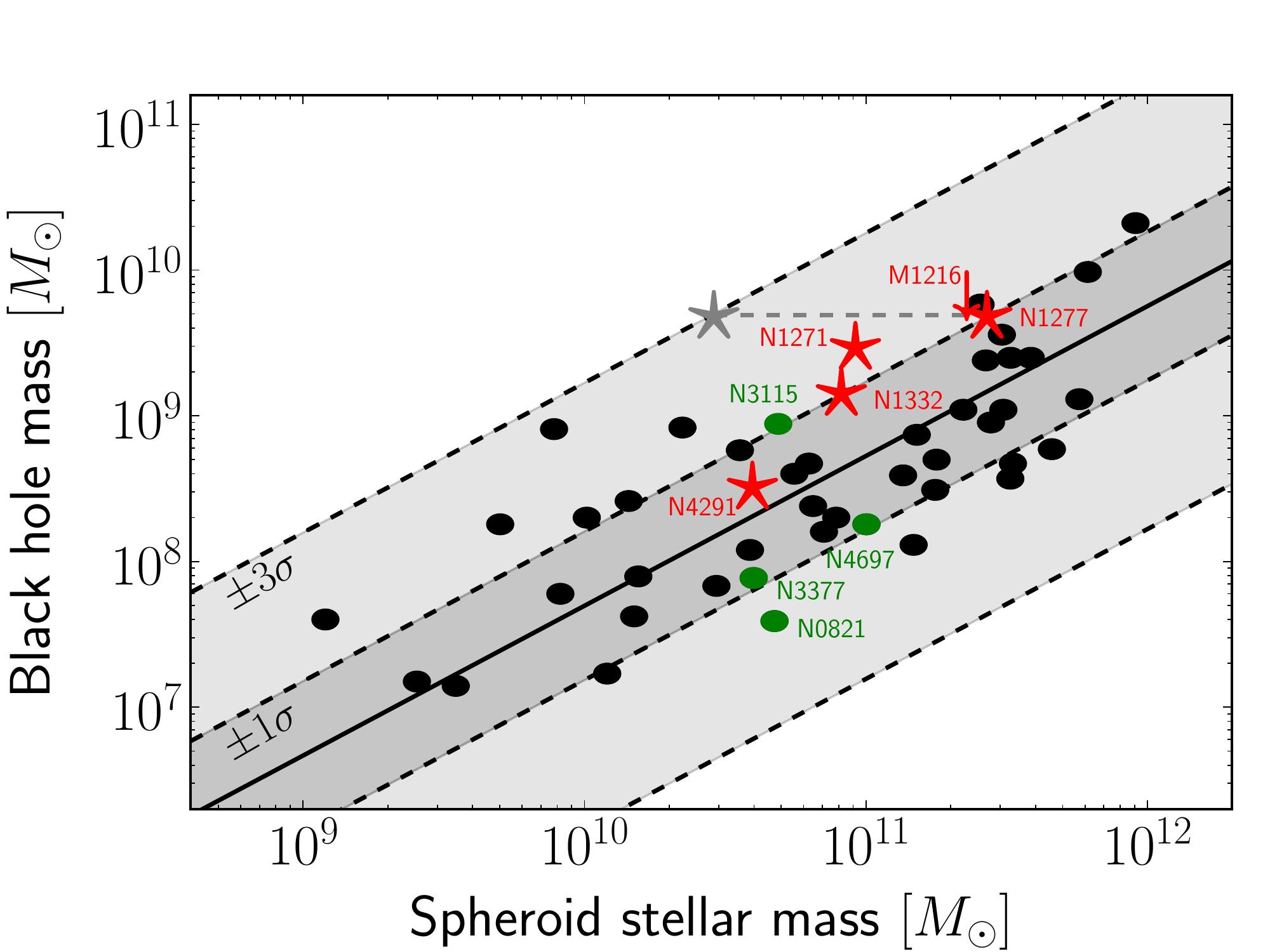}
\caption{Black hole mass plotted against spheroid stellar mass for 45+3 early-type galaxies (from \citealt{paperII}). 
The black solid line is the bisector linear regression for all galaxies except Mrk 1216, NGC 1271 and NGC 1277. 
The dashed lines mark the $1\sigma$ and $3\sigma$ deviations, 
where $\sigma$ ($0.51$ dex) is the total \emph{rms} scatter about the correlation in the black hole mass direction. 
The red symbols mark five galaxies that were claimed to be extreme outliers in this diagram: 
four intermediate-scale disc galaxies (Mrk 1216, NGC 1271, NGC 1277 and NGC 1332) and one elliptical galaxy (NGC 4291). 
All five reside well within a $3\sigma$ deviation from the correlation when using their correct spheroid mass. 
For NGC 1277, we show the previously reported spheroid stellar mass \citep{vandenbosch2012} in gray. 
The green color is used to show the location of four additional intermediate-scale disc galaxies mentioned in Section \ref{sec:gal}.}
\label{fig:mm}
\end{center}
\end{figure}

Inaccurate measurements of the spheroid-to-total ratio of galaxies can impact galaxy scaling relations. 
Recently, a handful of galaxies with intermediate-scale discs have been claimed to host \emph{over-massive} black holes, 
i.e.~the mass of their central supermassive black hole has been reported to be significantly larger 
than what is expected from the galaxy's spheroid luminosity (or stellar mass).
This is the case for the galaxies Mrk 1216 (for which only an upper limit on its black hole mass has been published, 
\citealt{yildirim2015}), NGC 1271 \citep{walsh2015}, 
NGC 1277 \citep{vandenbosch2012,yildirim2015,walsh2015n1277} and NGC 1332 \citep{rusli2011}.
In addition to these, the elliptical galaxy NGC 4291 has also been claimed to be a $\sim$$3.6\sigma$ outlier 
above the (black hole mass)-(spheroid mass) scaling relation \citep{bogdan2012}. 
Obviously, having both the black hole mass and the spheroid mass correct is important 
for placing systems in the (black hole mass)-(spheroid mass) diagram. \\
\emph{At present, for early-type galaxies, the spheroid luminosity and the galaxy luminosity 
can be used to predict the black hole mass with the same level of accuracy\footnote{Note that 
\cite{lasker2014anal} reported that the spheroid luminosity and the galaxy luminosity are equally good tracers of the black hole mass 
irrespective of the galaxy morphological type, but their sample of 35 galaxies contained only 4 spiral galaxies. 
However, using a sample of 45 early-type and 17 spiral galaxies, 
\cite{paperII} shows that, when considering all galaxies irrespective of their morphological type, 
the correlation of the black hole mass with the spheroid luminosity is better than that with the galaxy luminosity.} 
\citep{paperII}. 
If a galaxy hosts a black hole that is over-massive compared to expectations from the spheroid luminosity, 
but whose mass is normal compared to expectations from the galaxy luminosity, 
one should wonder whether the spheroid luminosity might have been underestimated 
due to an inaccurate spheroid/disc decomposition. }
Indeed, none of the five galaxies just mentioned (Mrk 1216, NGC 1271, NGC 1277, NGC 1332, and NGC 4291) is a noticeable outlier 
in the (black hole mass)-(galaxy luminosity) diagram. 
In Figure \ref{fig:mm} we show the location of these five galaxies in the updated (black hole mass)-(spheroid stellar mass) diagram 
for early-type galaxies from \cite{paperII}. 
Figure \ref{fig:mm} was populated using the galaxy decomposition technique shown here 
and extensively described in \cite{paperI}. 
Briefly, we obtained Spitzer/IRAC $3.6~\rm \mu m$ images for 45 early-type galaxies 
which already had a dynamical detection of their black hole mass. 
We modelled their one-dimensional surface brightness profiles with a combination of analytic functions, 
using one function per galaxy component. 
Spheroid luminosities were converted into stellar masses using individual, 
but almost constant mass-to-light ratios ($\sim$$0.6$, \citealt{meidt2014}). \\
In Figure \ref{fig:mm}, we show the galaxies Mrk 1216, NGC 1271 and NGC 1277, 
which were not a part of the original sample of 45 early-type galaxies.
For the galaxy NGC 1271, we use the black hole mass measurement 
and the stellar mass-to-light ratio obtained by \cite{walsh2015}. 
For the galaxy NGC 1277, we use the black hole mass measurement obtained by \cite{walsh2015n1277} 
and the stellar mass-to-light ratio obtained by \cite{martinnavarro2015}. 
Note that for NGC 1277 we recover a spheroid stellar mass of $2.7 \times 10^{11}~\rm M_\odot$, 
in agreement with the value of $\approx 1-2 \times 10^{11}~\rm M_\odot$ obtained by \cite{emsellem2013} 
from his multi-Gaussian expansion models\footnote{In \cite{emsellem2013}, 
readers will find a clever discussion of the problematics asociated with the definition and the identification 
of the ``bulge'' component in a galaxy. }.  
For the galaxy Mrk 1216, we use the upper limit on the black hole mass 
and the stellar mass-to-light ratio obtained by \cite{yildirim2015}. 
For the first time, Figure \ref{fig:mm} reveals that when the four intermediate-scale disc galaxies Mrk 1216, NGC 1271, NGC 1277, NGC 1332, 
and the elliptical galaxy NGC 4291 are properly modelled, 
they no longer appear as extreme outliers above the (black hole mass)-(spheroid stellar mass) correlation for early-type galaxies, 
i.e.~they all reside well within a $3\sigma$ deviation from the correlation.

\section{Origin of compact massive galaxies}
\label{sec:disc}

Acknowledging the correct structure of galaxies with intermediate-scale discs is important to properly understand their origin. 
According to the current paradigm of cosmological structure evolution, 
the genesis of massive early-type galaxies is characterized by two distinct phases: ``in-situ'' and ``ex-situ''. 
The first phase takes place in a young Universe (within its first $4~\rm Gyr$), 
when cold gas inflows produced short and intense bursts of star formation 
that created compact and dense conglomerates of stars with high velocity dispersion (e.g.~\citealt{prieto2013}). 
These naked and compact conglomerates, named ``red nuggets'' \citep{damjanov2009}, 
have been observed at high-redshift with half-light sizes of 1--2 kpc \citep{daddi2005,trujillo2006,vandokkum2008}.
In the second phase (last $10~\rm Gyr$), discs and stellar envelopes 
were accreted around these primordial conglomerates and the external parts of today's galaxies assembled on scales of 2--20 kpc 
(e.g.~\citealt{driver2013}). \\
Today's Universe is populated by an abundance of compact, massive spheroids, 
with the same physical properties -- mass and compactness -- as the high-redshift red nuggets \citep{GDS2015}. 
Some of these local compact massive spheroids are encased within a large-scale disc, 
that is to say they are the bulges of some lenticular and spiral galaxies.  
Over the last $10~\rm Gyr$ their spheroids have evolved by growing a relatively flat disc (e.g.~\citealt{pichon2011,danovich2012,stewart2013})
-- rather than a three-dimensional envelope -- 
which has increased the galaxy size but preserved the bulge compactness. 
Of course some lenticular/ES galaxies may have been built from mergers (e.g.~\citealt{querejeta2015}, and references therein). 
The other compact massive spheroids of today's Universe belong to some galaxies with intermediate-scale discs. 
Indeed, Mrk 1216, NGC 1271, NGC 1277, NGC 1332, and NGC 3115 are all local compact intermediate-scale disc galaxies 
with purely old ($>10~\rm Gyr$) stellar populations. 
These galaxies have undergone the lowest degree of disc growth. \\
In addition to the observational clues as to the actual physical components in galaxies with intermediate-scale discs, 
one can reason on other grounds as to why these compact galaxies are not comprised of an inner bulge 
plus large-scale disc plus outer envelope. 
If they were such three-component systems, then one would have two possibilities. 
The first possibility is that these galaxies were already fully assembled $10~\rm Gyr$ ago; 
this would explain their old stellar populations, 
but it would also imply that their discs and envelopes had already formed during the first $4~\rm Gyr$ of the Universe, 
in disagreement with the current cosmological picture. 
The second possibility is that only their inner bulges (with sizes of 0.1--0.2 kpc, 
according to past decompositions) originated in the first $4~\rm Gyr$ 
and they subsequently accreted a substantial disc and envelope. 
If this was correct, then we would observe high-redshift, star-like, naked bulges with stellar masses 
within a factor of a few times the currently observed red nuggets but sizes which are $10$ times smaller. 
However, a dramatically different expectation is reached 
if one considers these galaxies today as spheroid-dominated systems with an intermediate-scale disc; 
in this case, both the galaxy size and the spheroid size are compact (1--2 kpc). 
This implies that, among the local descendants of the high-redshift red nuggets, 
the compact intermediate-scale disc galaxies have undergone the lowest degree of disc growth. 
That is, the bulk of a compact intermediate-scale disc galaxy quickly assembled ``in-situ'' in a very young Universe 
and experienced very little evolution over the last $10~\rm Gyr$.

\section{Summary and conclusions}
Early-type galaxies display a broad distribution of spheroid-to-total flux ratios (e.g.~\citealt{cappellari2011kmdr}), 
going from disc-less, ``pure'' elliptical galaxies (slow rotators) 
to disc-dominated lenticular galaxies (central fast rotators that continue to be fast rotating also beyond one half-light radius). 
In between these two extremes lie galaxies with intermediate-scale discs 
(spheroid dominated central fast rotators that become slow rotating in their outer regions), 
i.e.~discs of kiloparsec-size that remain ``embedded'' within the spheroidal component of the galaxy 
and do not dominate the galaxy light at large radii as large-scale discs do. 
While this is likely known to some readers, 
the surge of papers presenting galaxy decompositions which are not aware of this reality 
has created a pressing need for this reminder. 
We have shown that the light distribution of galaxies with intermediate-scale discs can be accurately described 
with a simple spheroid + disc (+ optional nuclear component) model, 
without the need for the addition of a bright envelope-component. \\
Our decompositions correctly reproduce both the photometric (surface brightness and ellipticity profiles) 
and kinematic (specific angular momentum profile) properties of nine intermediate-scale disc galaxies. 
Four of these nine galaxies (Mrk 1216, NGC 1271, NGC 1277, NGC 1332) and one additional elliptical galaxy (NGC 4291) 
had previously been claimed to be extreme outliers in the (black hole mass)-(spheroid mass) diagram. 
However, here we have demonstrated that, when correctly modelled, 
these five galaxies all reside well within the scatter of the correlation, 
i.e. they do not host over-massive black holes. 
This serves to strengthen the (black hole mass)-(spheroid mass) relation, 
and rules out the need for exotic formation scenarios. 

\section{Acknowledgments}
This research was supported by Australian Research Council funding through grant FT110100263. 
GS is grateful to Matteo Fossati, Luca Cortese and Giuseppe Gavazzi for useful comments and discussion. 
The publication of this paper would not have been possible without the invaluable support of 
Chris Blake and Duncan Forbes. 
We warmly thank our anynimous referee for their very careful review of our paper, 
and for the comments, corrections and suggestions that ensued. 
This work is based on observations made with the IRAC instrument \citep{fazio2004IRAC} on-board the Spitzer Space Telescope, 
which is operated by the Jet Propulsion Laboratory, California Institute of Technology under a contract with NASA, 
and also on observations made with the NASA/ESA Hubble Space Telescope, 
and obtained from the Hubble Legacy Archive, 
which is a collaboration between the Space Telescope Science Institute (STScI/NASA), 
the Space Telescope European Coordinating Facility (ST-ECF/ESA) and the Canadian Astronomy Data Centre (CADC/NRC/CSA).
This research has made use of the GOLDMine database \citep{goldmine} and the NASA/IPAC Extragalactic Database (NED) 
which is operated by the Jet Propulsion Laboratory, California Institute of Technology, 
under contract with the National Aeronautics and Space Administration. 

\bibliography{SMBHbibliography}

\label{lastpage}

\clearpage

\end{document}